# The benefits of a Bayesian analysis for the characterization of magnetic nanoparticles

*Dr. Mathias Bersweiler**
*Department of Physics and Materials Science, University of Luxembourg, L-1511, Grand Duchy of Luxembourg*
*E-mail:mathias.bersweiler@uni.lu*

*Dr. Helena Gavilan Rubio*
*a) Istituto Italiano di Technologia, Central Research Labs Genova, Via Morego 30, 16163 Genova, Italy*
*b) Institute of Material Science of Madrid, ICMM/CSIC, C/Sor Juana Inés de la Cruz 3, 28049 Madrid, Spain*

*Dr. Dirk Honecker*
*Department of Physics and Materials Science, University of Luxembourg, L-1511, Grand Duchy of Luxembourg*

*Prof. Dr. Andreas Michels**
*Department of Physics and Materials Science, University of Luxembourg, L-1511, Grand Duchy of Luxembourg*
*E-mail: andreas.michels@uni.lu*

*Dr. Philipp Bender**
*Department of Physics and Materials Science, University of Luxembourg, L-1511, Grand Duchy of Luxembourg*
*E-mail:philipp.bender@uni.lu*

*Abstract:* Magnetic nanoparticles offer a unique potential for various biomedical applications, but prior to commercial usage a standardized characterization of their structural and magnetic properties is required. For a thorough characterization, the combination of conventional magnetometry and advanced scattering techniques has shown great potential. In the present work, we characterize a powder sample of high-quality iron oxide nanoparticles that are surrounded with a homogeneous thick silica shell by DC magnetometry and magnetic small-angle neutron scattering (SANS). To retrieve the particle parameters such as their size distribution and saturation magnetization from the data, we apply standard model fits of individual data sets as well as global fits of multiple curves, including a combination of the magnetometry and SANS measurements. We show that by combining a standard least-squares fit with a subsequent Bayesian approach for the data refinement, the probability distributions of the model parameters and their cross correlations can be readily extracted, which enables a direct visual feedback regarding the quality of the fit. This prevents an overfitting of data in case of highly correlated parameters and renders the Bayesian method as an ideal component for a standardized data analysis of magnetic nanoparticle samples.

## 1. Introduction

Magnetic nanoparticles have been intensively studied in recent years partly because they are promising candidates for various biomedical approaches [1–4]. However, prior to clinical application, the particle properties need to be characterized in a thorough and well established (*i.e.* standardized) way [5–7]. Structural parameters, such as size and shape, can be determined *e.g.* by transmission electron microscopy (TEM), dynamic light scattering (DLS), or small-angle x-ray scattering (SAXS). Magnetic properties on the other hand, such as the particle moment and saturation magnetization, can be evaluated with DC magnetometry (DCM), Mössbauer spectroscopy, small-angle neutron scattering (SANS), or X-ray resonant magnetic scattering (XRMS). Especially the combination of standard techniques with more advanced scattering methods enables a detailed characterization of magnetic nanoparticle ensembles. For instance, XRMS combined with TEM and DCM has been used (i) to reveal the nanoscale magnetic ordering in self-assembled particles [8] or (ii) to model inter-particle magnetic correlations [9]. On the other hand, SANS combined with TEM and DCM has been employed to investigate the dipolar-coupled moment correlations in nanoparticle ensembles [10], the in situ dimensional characterization of magnetic nanoparticle clusters during induction heating [11] or to elucidate the interplay between the particle size and the magnetization configuration [12–14].

To retrieve particle properties from experimental data, usually the data sets are fitted with analytical models, *e.g.* the magnetization curve of magnetic colloids with the Langevin function [15]. However, in most of the studies, a least-squares algorithm is implemented to minimize the sum of square residuals between the measured data and the analytical model. Algorithms such as the Levenberg-Marquardt method [16,17] can quickly converge to a solution, if the starting parameters are close to the global minimum. But in case of non-linear models with complex functions and/or with many parameters, the solution can be trapped into local minima or the algorithm may not converge. The obtained fitting parameters thus do not necessarily reflect the *real* values and the obtained results using least-squares algorithms must be interpreted carefully.

An alternative approach to derive the model parameters from model fits is by Bayesian inference. The main advantage of the Bayesian approach is that the complete probability distributions of the model parameters and their cross correlations are sampled. A plot of the distributions and their cross correlations offers a direct visual feedback regarding the quality of the fit by which *e.g.* an overfitting of the data can be immediately spotted. A brief summary of the substantial advantages of the Bayesian approach compared to the standard fitting approach which are generally employed in most of the studies can be found in Ref. [18]. The benefit of this approach has already been demonstrated for the analysis of Rietveld refinement of X-ray and neutron diffraction patterns [19], the structure analysis of macromolecules by small angle scattering (SAS) [20] as well as for the estimation of the particle size or the correlation lengths of colloidal nanocrystals by SANS [21,22]. More recently, the Bayesian approach was also used to optimize the refinement of neutron reflectometry data [23].

In this study we exemplarily analyze the DCM and magnetic SANS data of a powder sample of high-quality iron oxide nanoparticles that are coated with a thick silica shell. We apply the Bayesian approach for the refinement of the data to retrieve the structural and magnetic parameters of the particle ensemble, and we discuss our findings and the general strengths of this method.

## 2. Methods

The spherical iron-oxide nanoparticles (IONPs) were synthesized by a high-temperature thermal decomposition method [24]. The magnetite core particles were then coated with a silica shell by a reverse microemulsion method [25]. In the following, the total particle size (*i.e.* the outer shell size) and the inner core size will be defined as $D_s = 2R_s$ and $D_c = 2R_c$, respectively. For the DCM and magnetic SANS measurements, we used a dense powder sample of the IONPs. Thanks to the silica shell, dipolar interactions between the IONPs are significantly reduced [26,27].

The main structural properties of the IONPs were estimated by transmission electron microscopy (TEM) using a JEOL microscope (Peabody, USA) operating at 100 kV. TEM samples were prepared by placing a drop of the particles suspended in water onto a carbon-coated copper grid and allowing it to dry at room temperature. The size distribution was determined by manual measurement of more than 100 particles using the open-access ImageJ software [28].

The magnetic analysis of the particle powder at room temperature was performed using a commercial (Quantum Design) superconducting quantum interference device (SQUID) magnetometer. The diamagnetic contribution of the sample holder was subtracted. Moreover, the measured moment was normalized to the iron content of the sample (which was detected by inductively-coupled plasma optical emissions spectrometry, ICP-OES [7]) so that we obtained the field-dependent magnetization curve $M(H)$ in units of $Am^2/kg_{Fe}$.

The neutron experiments were conducted at the instrument D33 at the Institut Laue-Langevin, Grenoble, France [29]. As sample holder for the particle powder, we used a standard quartz-glass cuvette with a thickness of 1 mm. The measurements were realized using a polarized incident neutron beam with a mean wavelength of $\lambda = 6$ Å and a wavelength broadening of $\Delta\lambda/\lambda = 10\%$ (FWHM). An external magnetic field $\boldsymbol{H}_0$ (up to 5T) was applied perpendicular to the incident beam ($\boldsymbol{H_0} \parallel \boldsymbol{e}_z \perp \boldsymbol{k_0}$). The measurements were performed at room temperature within a $q$-range of 0.16 nm$^{-1}$ ≤ $q$ ≤ 0.5 nm$^{-1}$. For the neutron-data reduction the GRASP software package was used [30]. By employing the so-called SANSPOL analysis, detailed *e.g.* in Ref. [31–33], the half-polarized cross-sections d$\Sigma^+$/d$\Omega$ and d$\Sigma^-$/d$\Omega$ were measured. From these measurements, we determined the nuclear-magnetic cross term $I_{\text{cross}}(q)$ by subtraction of the two half-polarized SANSPOL cross sections, *i.e.* d$\Sigma^-$/d$\Omega$ − d$\Sigma^+$/d$\Omega$. The analysis of the spin-state difference has the advantage that background-scattering contributions, such as incoherent scattering, are absent. The intensity in the limit of $q \to 0$ is proportional to the magnetic moment of the particle ensemble and can be directly compared to macroscopic magnetization measurements. The method was used *e.g.* to characterize magnetic core-shell particles [34], the magnetic profiles [35] and the spatial magnetization distribution in magnetic nanoparticles [13,36], the relaxation mechanisms in magnetic colloids [37], or more recently to investigate the response of magnetotactic bacteria under a magnetic field [38]. Here, we analyzed the cross term $I_{\text{cross}}(q)$ of the IONP powder sample measured at different field strengths, namely at $\mu_0 H_0$ = 0.02, 0.1, 1.0 and 5 T, to derive the structural properties (such as $D_s$ and $D_c$).

## 3. Analytical models

To extract the particle parameters from the DCM measurement $M(H)$ and the nuclear-magnetic cross term $I_{\text{cross}}(q)$, we fitted the data sets with the analytical models presented below. In doing so, we assumed a spherical shape and that the iron-oxide cores were homogeneously magnetized. This means that (i) the magnetic, the core, and the total particle volumes are given by: $V_i(R_i) = 4\pi R_i^3/3$ with $R_i$ the magnetic ($i = m$), the core ($i = c$), and the outer shell ($i = s$) radius. (ii) The magnetic particle moment is taken as $\mu(R_m) = M_s V_m(R_m)$, where $M_s$ is the saturation magnetization of the material. (iii) The magnetic scattering amplitude $F_{mag}(q, R_c)$ is proportional to the nuclear one of the iron oxide core (*i.e.* $R_m = R_c$).

### 3.1 DCM data

In case of superparamagnetic nanoparticles, the DCM curve follows the Langevin function. For polydisperse samples, additionally a particle-size or moment distribution needs to be considered. By applying numerical inversion methods these distributions can be extracted without any prior assumption regarding the particle size [15,39–47]. However, usually the shape of the particle-size distribution of MNPs is assumed *a priori* (*e.g.* a Gaussian or log-normal shape [48,49]). Moreover, assuming a constant magnetic moment of the particle, the magnetization behavior of non-interacting MNP ensembles can be modeled by:

$$M(H) = M_S \int_0^\infty \frac{P(R_m) R_m^3 L(H, \mu(R_m)) dR_m}{P(R_m) R_m^3 dR_m} + b\mu_0 H \quad (1)$$

with $R_m$ being the magnetic particle radius (in our case we can assume that $R_m = R_c$) and $L(H, \mu(R_m)) = \coth\left(\frac{\mu(R_m)\mu_0 H}{k_B T}\right) - \frac{k_B T}{\mu(R_m)\mu_0 H}$ being the Langevin function. Here, $\mu_0$ is the vacuum permeability, $k_B$ is the Boltzmann constant and $T$ is the absolute temperature. In the following, we will assume that the magnetic particle-size distribution $P(R_m)$ follows the log-normal distribution:

$$P(R_m) = \frac{1}{\sqrt{2\pi}\sigma_m R_m} e^{\frac{-\ln(R_m/\tilde{R}_m)^2}{2\sigma_m^2}} \quad (2)$$

with $\sigma_m$ being the standard deviation of the log-normal distribution and $\tilde{R}_m$ being the median value of the magnetic radius. The second term $b\mu_0 H$ in equation 1 accounts for linear magnetization contributions (usually paramagnetic contributions), which may originate *e.g.* from uncorrelated surface spins [50] and canted sublattice spins [51].

### 3.2 Magnetic SANS data

For polydisperse core-shell particle ensembles, the nuclear-magnetic cross term $I_{\text{cross}}(q)$ can be written as [13]:

$$I_{\text{cross}}(q) = \iint F_{nuc}(q, R_c, R_s) F_{mag}(q, R_c) P(R_c) P(R_s) dR_c dR_s \quad (3)$$

with $F_{nuc}(q, R_c, R_s)$ and $F_{mag}(q, R_c)$ being the nuclear and magnetic scattering amplitudes of the spherical (core-shell) particles, respectively. Both contributions are defined as [52]:

$$F_{nuc}(q, R_c, R_s) = \frac{3V_c}{qR_c}(\rho_c - \rho_s)j_1(qR_c) + \frac{3V_s}{qR_s}\rho_s j_1(qR_s) \quad (4)$$

$$F_{mag}(q, R_c) = \frac{3V_c}{qR_c}\rho_m j_1(qR_c) \quad (5)$$

Here, $j_1(qR_i) = \sin(qR_i)/(qR_i)^2 - \cos(qR_i)/(qR_i)$ is the first-order spherical Bessel function, and $\rho_i$ is the nuclear scattering length density of the core ($i = c$) and the outer shell ($i = s$), respectively. $\rho_m$ is the magnetic scattering length density of the magnetic core, which is directly proportional to the sample magnetization. In case of superparamagnetic nanoparticle ensembles, $\rho_m$ can thus be theoretically replaced by the Langevin function [13].

### 3.3 Bayesian approach

The Bayesian refinement of the magnetometry and magnetic SANS data was performed using the Bayesian algorithm implemented in the *emcee* open-source Python package developed by Foreman-Mackey *et al.* [53] and which is based on the J. Goodman and J. Weare's affine invariant Markov chain Monte Carlo (MCMC) Ensemble samplers [54]. Briefly, the *emcee* open-source package returns the log-posterior probability $\ln p(\boldsymbol{P}|\boldsymbol{D})$ of a set of model parameters $\boldsymbol{P}$ given the data $\boldsymbol{D}$. This log-posterior probability is determined from the Bayes' theorem and is proportional to:

$$\ln p(\boldsymbol{P}|\boldsymbol{D}) \propto \ln p(\boldsymbol{D}|\boldsymbol{P}) + \ln p(\boldsymbol{P}) \quad (6)$$

with $\ln p(\boldsymbol{P})$ being the log-prior knowledge of the model parameters $\boldsymbol{P}$ and $\ln p(\boldsymbol{D}|\boldsymbol{P})$ being the log-likelihood function (probability of observing the data $\boldsymbol{D}$ given the set of model parameters $\boldsymbol{P}$). Experimentally, equation 6 is determined by minimizing the negative log-likelihood function, which is defined in the *emcee* package as:

$$\ln p(\boldsymbol{D}|\boldsymbol{P}) = -\frac{1}{2}\sum_i \left[\frac{(f_i(\boldsymbol{P}) - D_i)^2}{\sigma_i^2} + \ln(2\pi\sigma_i^2)\right] \quad (7)$$

with $f_i$ being the generative model, $\sigma_i$ the measurement uncertainty, and $\boldsymbol{D}_i$ the collected data at a given point "$i$". For more details about Bayesian theory, we refer to the book by D'Agostini [55].

To calculate the log-posterior probability distribution of the parameters given a set of experimental data, we use the function *minimize* implemented in the *lmfit* open-source Python package developed by Newville *et al.* [56]. As starting values for the Bayesian refinement of the free fit parameters we use the results of a prior standard least-squares fit using the Levenberg-Marquardt algorithm. From the obtained probability distributions, the median values plus the standard deviation $\sigma$ (*i.e.* half the difference between the 15.8 and 84.2 percentiles of the distribution) are used to define the *best-fitting* parameters.

## 4. Results and discussion

### 4.1. TEM analysis

Figure 1(a) shows a representative TEM image of the IONPs. As can be seen, the dark iron oxide cores are nearly perfect spheres that are surrounded by a brighter, homogenous layer of silica. The derived histograms of the total particle size (*i.e.* the outer shell diameter) and the core size are displayed in figures 1(b) and (c), respectively. These histograms were fitted with a log-normal distribution (in this case *via* a least-squares fit), and the best-fitting parameters are summarized in Table 1. Overall, TEM confirms that the particles have a low polydispersity. The mean core size is $\langle D_c \rangle = 12.56 \pm 0.03$ nm and the broadness is $\sigma_c = 0.08 \pm 0.01$. The mean total particle size is $\langle D_s \rangle = 42.5 \pm 0.2$ nm with $\sigma_s = 0.12 \pm 0.01$. These results will be used as a reference for the fit of the DCM and magnetic SANS data.

### 4.2. DCM analysis

Figure 2(a) displays the room temperature magnetization curve of the IONP powder sample. The measured curve shows no hysteresis, indicating a superparamagnetic behavior of the iron oxide cores. This is expected for IONPs (in this case maghemite particles) with diameters below about 25 nm [57]. We fitted the magnetization branch $\mu_0 H = +5 \to -5 T$ with equation 1 using the Bayesian approach. All fitting parameters were allowed to vary and were unconstrained (see the supporting information for the open-source python code). As starting values, we used the results of the standard least-squares fit using the Levenberg-Marquardt algorithm.

The best fit is shown in figure 2(a) (red line) and the corresponding residuals are plotted in figure 2(b). Figure 2(c) shows the probability distributions of each fitting parameter and the two-dimensional plots of their respective cross correlations. The histogram of each individual parameter exhibits a narrow and symmetric peak (*i.e.* a normal (Gaussian) distribution) which indicates that the solution found from the Bayesian refinement is a strong minimum of the $\chi^2$ function [19]. The median values of these distributions (*i.e.* the *best*-fit values) are summarized in Table 1. The determined value for the saturation magnetization $M_S$ = 95.5 ± 0.3 Am²/kg$_{Fe}$ is similar to the values reported in the literature for superparamagnetic IONPs [58]. The values for the log-normal distribution of the magnetic particle size (*i.e.* the width $\sigma_m$ and the median value $D_m$) are in good agreement with the values $\langle D_c \rangle$ and $\sigma_c$ determined by TEM for the iron oxide cores. However, the size distribution extracted from the magnetization curve is slightly broader compared to TEM (figure 2(d)). This could be attributed to a slight deviation from the Langevin-like magnetization behavior. Possible reasons for this may be (i) dipolar interactions between the particles *e.g* due to some aggregated iron oxide cores, and/or (ii) a finite contribution of the magnetocrystalline anisotropy [59]. Such deviations could also explain the small but systematic oscillation of the residuals around zero, which indicates that the fit model cannot describe the experimental data perfectly.

Regarding the cross-correlation plots in figure 2(c), an isotropic pattern in the center of the box indicates that there is no strong correlation between the values of two parameters [19,22]. This is the case for example for the pairs ($M_S, \sigma_m$) and ($b, \sigma_m$); they are essentially independent of each other. On the other hand, the observation of an elongated ellipsoid pattern results from a strong correlation between two parameters. The sign of the correlation defines the slope of the ellipsoid. This particular pattern shape is clearly seen for the pairs ($\sigma_m, D_m$) and ($b, M_S$); this means that in both cases the two parameters are strongly correlated [19,22].

Considering the strong correlation between ($\sigma_m, D_m$), the before mentioned broader size distribution determined from the fit of the DCM data compared to TEM (Figure 2(d)) is not unambiguous. In fact, when fixing $\sigma_m$ to the TEM value ($\sigma_m = \sigma_c = 0.08$), the total $\chi^2$ is only slightly larger compared to the *best* fit before (compare residuals in figure 2), and we obtain $D_m = 12.44 \pm 0.05$ nm. This is basically identical to the TEM value and, consequently, we get for the (magnetic) core-size distribution practically the same one as for TEM (see Figure 2(d)). This nicely shows the ambiguity of model fits in case of highly correlated parameters, as it is commonly the case for distribution functions. But these correlations can be easily detected by the cross-correlation plots with the Bayesian approach.

**4.3. Magnetic SANS analysis**

In this section, the Bayesian methodology is applied for the refinement of the magnetic SANSPOL data, *i.e.* the nuclear-magnetic cross term $I_{cross}(q)$. The Bayesian refinement was performed on a single data set and then simultaneously on the multiple, field-dependent data sets with some shared (*i.e.* global) fit parameters. In both cases, we systematically reduced the number of free fitting parameters to prevent an over-fitting of the data, as discussed below.

*4.3.1. Single data set*

Figure 3(a) displays the nuclear-magnetic cross term $I_{cross}(q)$ at a magnetic field of $\mu_0H_0 = 5$T, which is close to the magnetic saturation of the IONPs (see magnetization curve in figure 2). We fitted $I_{cross}(q)$ with equation 3 using the Bayesian approach. For the nuclear scattering length density of the core we assumed $\rho_c = 6.9 \times 10^{-6}$ Å$^{-2}$ which is the expected value for maghemite (Fe$_2$O$_3$) in case of a density of 5.0 g/cm³.

First, all 6 parameters (*i.e.* $D_c, D_s, \sigma_c, \sigma_s, \rho_s, \rho_m$) were allowed to vary. As can be seen in figures 3(a) and (b), the fit (green line) describes at first sight the experimental data well. The best-fitting parameters are summarized in Table 2, and the corresponding log-normal size distributions of the outer shell and the iron oxide core are displayed in figures 3(c) and (d). It can be seen that the distribution of the total particle size is in quite good agreement with the TEM result albeit slightly broader. The distribution of the core size, however, is unphysically narrow. The probability distribution of each fitting parameter and their respective cross correlations in figure 4(a) provide additional information. The extremely skewed probability distributions for some of the parameters (especially $\sigma_c$) and the irregular shape observed for many of the cross correlations (*e.g.* between $D_c$ and $\sigma_c$) confirm that the fit with all 6 fit parameters being free and unconstrained does overfit the data.

To reduce the number of free parameters, we fixed $\rho_s$ to 2.8 ×10$^{-6}$ Å$^{-2}$. This is the expected value for the nuclear scattering length density (SLD) of silica calculated using the software SasView (www.sasview.org) assuming a density of 1.8 g/cm³, as expected for silica nanoparticles according to Ref. [60]. The best fit with the remaining 5 free fit parameters, the corresponding residuals and the estimated log-normal distribution of the outer shell and the core size (computed with the best fitting parameters summarized in Table 2) are sketched by the blue lines in figure 3. Again, a qualitatively good agreement between fit and data is observed. But by reducing the number of free fit parameters, the probability distribution of each remaining parameter approaches a normal distribution, as expected (compare figures 4(a) and (b)). This indicates a more robust solution and thus a good fit. However, as shown in figure 3(d), the estimated log-normal distribution of the core size is significantly shifted to lower values compared to TEM. $\sigma_s$ and $D_c$ are negatively correlated (anisotropic cross correlation with negative slope, figure 4(b)). Thus a possible explanation for the shift in $D_c$ is an increased value $\sigma_s$ caused by a slight deviation of the total particle shape from perfect spheres (see TEM image in figure 1). As can be seen, the thickness of the silica layer around the cores is not perfectly homogeneous. When performing the fit with the spherical core-shell model, we then obtain a larger $\sigma_s$ but which is unphysical and simply a result of the erroneous model. This causes, due to the negative cross correlation between both parameters, a shift of the core size $D_c$ distribution to lower values. To circumvent this issue, we fixed the value of $\sigma_s$ to the TEM value of 0.12 and performed a fit with only 4 free fit parameters left. It can be noted, that *e.g.* also for the analysis of neutron reflectometry data [23] the maximum number of possible free fit parameters was found to be 4.

The results with the 4 remaining free fit parameters are displayed by the red lines in figure 3. The probability distribution of each fitting parameter and the cross correlations are shown in figure 4. In this case, the log-normal distribution of the core size (see best fit parameters in Table 2) is in very good agreement with the TEM histogram. However, the probability and the cross-correlation distributions seem to be slightly distorted, *i.e.* asymmetric. This indicates a certain ambiguity of the fit. To increase the confidence of the solution, we performed a global fit of all field-dependent measurements.

*4.3.2 Global fit of multiple data sets*

Figure 5(a) shows the plot of the nuclear-magnetic cross term measured at the 4 applied magnetic fields ($\mu_0 H_0$ = 0.02, 0.1, 1.0 and 5.T). As expected, the cross term significantly varies with the applied field. In particular, the magnitude of the scattering intensity increases with increasing fields, which can be attributed to an increased magnetization of the sample in the field direction. In our model (equation 3) this field-dependency is covered by the magnetic scattering length density $\rho_m$. Thus, when fitting the four data sets simultaneously, $\rho_m$ must be adjusted for each cross term individually. $\rho_s$, on the other hand, was, as before, fixed to 2.8 ×10$^{-6}$ Å$^{-2}$. Furthermore, we assumed that the 4 structural parameters (*i.e.* $D_c, D_s, \sigma_c, \sigma_s$) are field-independent and thus that they can be handled as global fit parameters.

As shown in figure 5, the fit with the 8 free fit parameters adjusts the field-dependent cross sections very well, and for all parameters we obtain symmetric, normal-shaped distributions (figure 6). However, as also observed for the fit of the single data set before, the distribution (computed with the best-fitting parameters summarized in Table 3) of the total particle size is again broader than the TEM result and the distribution of the core size is shifted to lower values. As before, this discrepancy can be explained by the negative correlation between these two parameters. Due to the slight deviation of the particle shape from perfect spheres caused by the slightly inhomogeneous thickness of the silica layer, the fit results in an artificially large $\sigma_s$ value, which in turn leads to a reduced value for the core size. To circumvent this issue, we fixed $\sigma_s$ to the TEM value $\sigma_s = 0.12$, as done before.

The blue lines in figure 5 display the fit result in case of the 7 remaining free fit parameters, and in figure 6, we show the corresponding distributions and cross correlations. In this case, we get an excellent agreement for the particle-size distributions in comparison with TEM. Furthermore, the results are basically identical to the results of the single data set using 4 free fit parameters (see Table 2). However, in contrast to the single data fit, now the probability and cross-correlation distributions of all free fit parameters (figure 6) are symmetric. This shows that by performing such a global fit the solution is more robust, which ultimately leads to an increased confidence regarding the fit result.

Finally, we performed a global fit of the SANS and magnetometry data. As mentioned above, the values for the field-dependent magnetic scattering length density $\rho_m$ should be directly proportional to the macroscopic magnetization of the sample. Therefore, we assumed that $\rho_m = cM_sV_m L(H, \mu(R_m))$, where $c$ is a scaling factor. In figure 7, we plot the normalized values of $\rho_m$ (yellow stars) obtained from the fit of the SANS data. It can be seen that they agree quite well with the measured isothermal magnetization curve at the three higher fields. At the lowest field, we observe a slight shift. We surmise that this shift is due to the remanence of the cryostat we used for the SANS experiment which would explain a reduction of the actual magnetic field compared to the nominal value of $\mu_0 H_0$ = 0.02 T. Thus, to perform a combined global fit of the magnetization and SANS data, we kept the applied field strength of the low-field SANS measurement as a free fit parameter. Furthermore, we had to fix $\rho_s$, $\sigma_s$ and $\sigma_c$, as otherwise the resulting parameter distributions were highly distorted which indicated an overfitting of the data (*i.e.* too many free and highly correlated fit parameters). The resulting fit shown in figure 8 is in good agreement with the TEM results regarding the mean particle sizes. Moreover, the isotropic pattern observed for the pair $(D_c, D_s)$ indicates uncorrelated parameters. This result, which differs from the global fit performed only on the SANS data (compare with figure 6(b)), indicates more accurate best-fitting parameters (see table 4) for the core and particle size by performing the global fit on the SANS+DCM datasets. This emphasizes the benefits of using global fits of complementary techniques for a characterization of magnetic nanoparticle samples. By using the Bayesian approach for this endeavor, the quality of the fit can be readily evaluated, which is the main strength of this method.

**5. Conclusion**

In this work we applied a Bayesian approach for the refinement of magnetization and magnetic SANS data of magnetic nanoparticles. As model system, we investigated spherical superparamagnetic iron oxide cores that are covered by a thick silica layer. We fitted the isothermal magnetization curve and the magnetic SANS data (to be precise, the nuclear-magnetic cross term) using standard analytical models discussed in section 3 convoluted with log-normal distributions to obtain the core and total particle size distributions. The novel aspect of this study is that we applied a Bayesian approach for the data refinement, which enabled us to obtain for each free fit parameter its probability distribution and the cross correlations for all parameter pairs. This method allows one to detect *e.g.* an overfitting of the data and enables the detection of critical correlations between the fit parameters. Considering that the Bayesian approach provides a direct visual feedback regarding the quality of the fit we thus propose the combination of a conventional least-squares fit with a subsequent Bayesian refinement as the new standard approach for data fitting of magnetic nanoparticle samples. Such a standardized protocol for the data refinement would be especially useful with biomedical applications in mind, as regulatory work regarding prior particle characterization is still required to guarantee a safe and effective administration of magnetic nanoparticles into the human body.

The computer code for the fit was written in python using open-source python packages (including the *emcee* and *lmfit* packages) and examples for the refinement can be found in the GitHub repository https://github.com/PBenderLux/Data-analysis.

**Acknowledgements**
The authors acknowledge the Institut Laue-Langevin for provision of neutron beamtime. P.B. and A.M. acknowledge financial support from the National Research Fund of Luxembourg (CORE SANS4NCC grant). The authors thank David González-Alonso, Imanol de Pedro and Luis Fernández Barquín for help with the magnetization and SANS measurements. SasView contains code developed with funding from the European Union's Horizon 2020 research and innovation programme under the SINE2020 project, grant agreement No 654000.

**Figure 1.** (a) Transmission electron microscopy (TEM) images of the iron oxide nanoparticles coated with silica. (b) and (c) Histograms ($N$ = 216 particles) of the total particle size and core size (denoted $D_s$ and $D_c$, respectively) fitted with a log-normal distribution function (red lines).

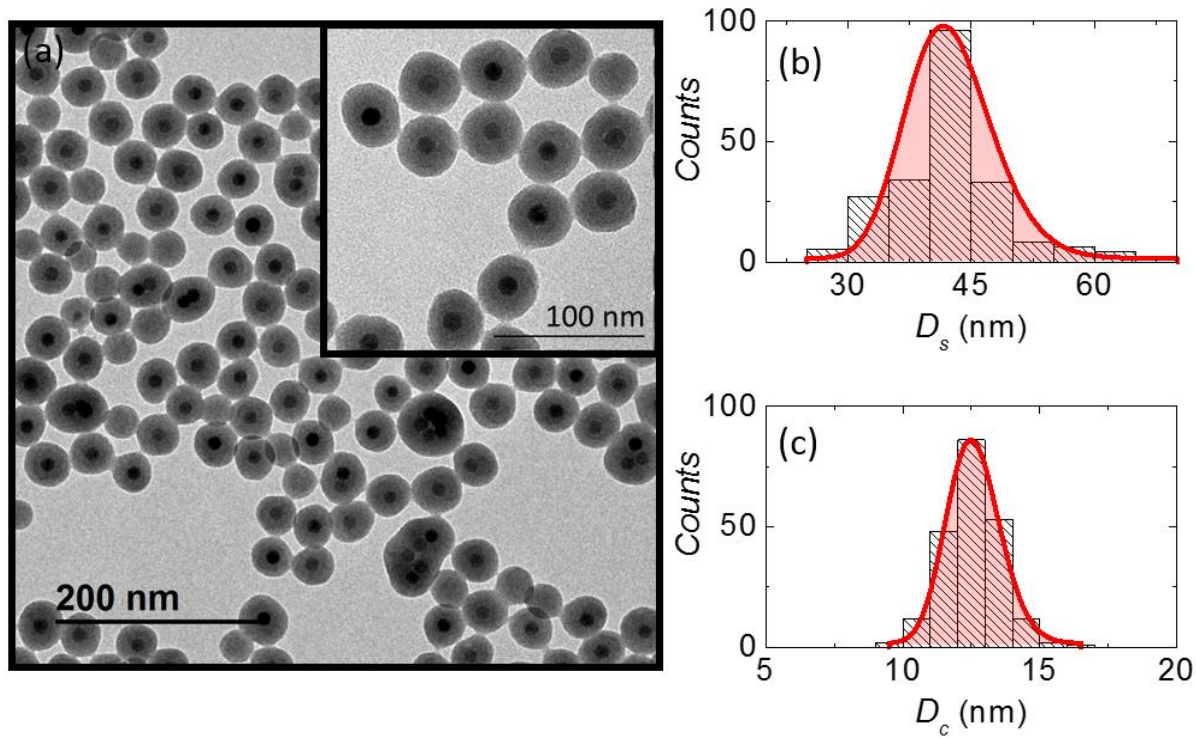

**Figure 2.** (a) Magnetization curve measured at room temperature in a field range of ± 5 T (open squares). Solid red line: best fit of the data by using the Bayesian algorithm and equation 1 as fitting model. (b) Corresponding residuals. Open symbols are the residuals for $\sigma_m = \sigma_c = 0.08$. (c) Corner plot of the fitting parameters obtained from the Bayesian refinement. The main diagonal displays the 1D probability distribution (histogram) for each fitting parameter. The off diagonal two-dimensional plots show the cross correlations for each pair of parameters. (d) Log-normal size distribution of the magnetic particles computed for the best-fitting parameters (red line). The dashed red line corresponds to the size distribution when $\sigma_m$ is fixed to the core-size value 0.08 determined by TEM. Units: $D_m$ in nm and the saturation magnetization $M_s$ in Am$^2$/kg$_{Fe}$. To estimate the volume magnetization in [A/m] (which is needed to transfer the particle moment to its size) we assumed the iron oxide density to be $\rho_{\gamma-Fe_2O_3} = 5000$ kg/m$^3$ and the iron content to be 70%.

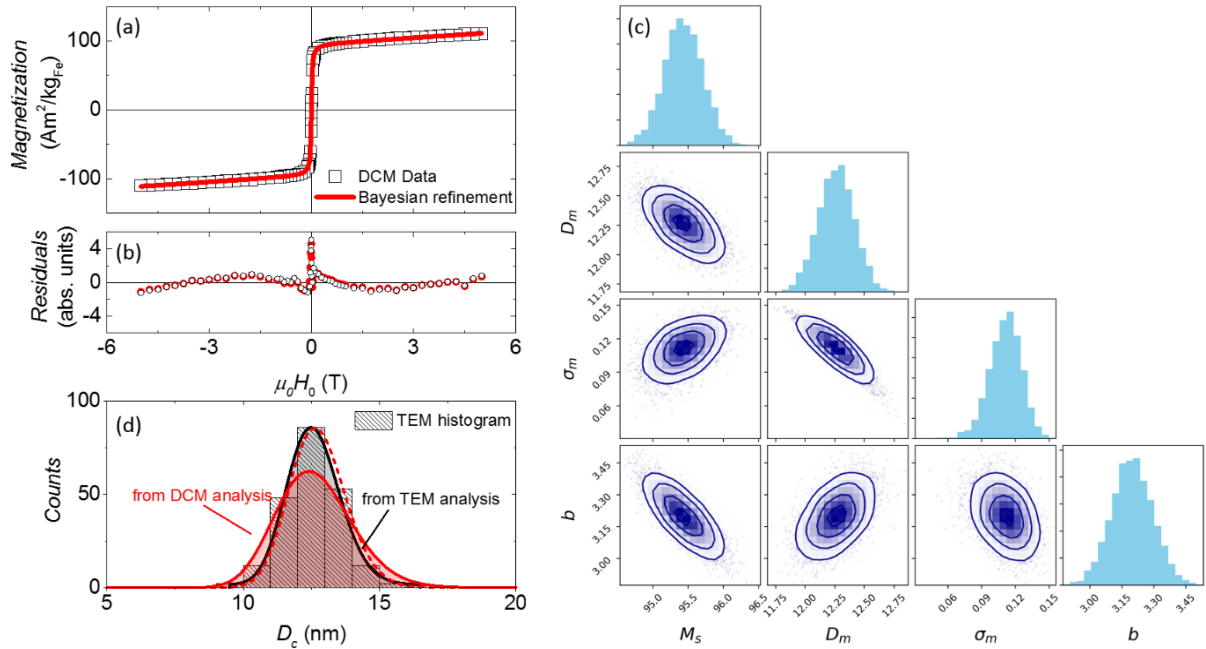

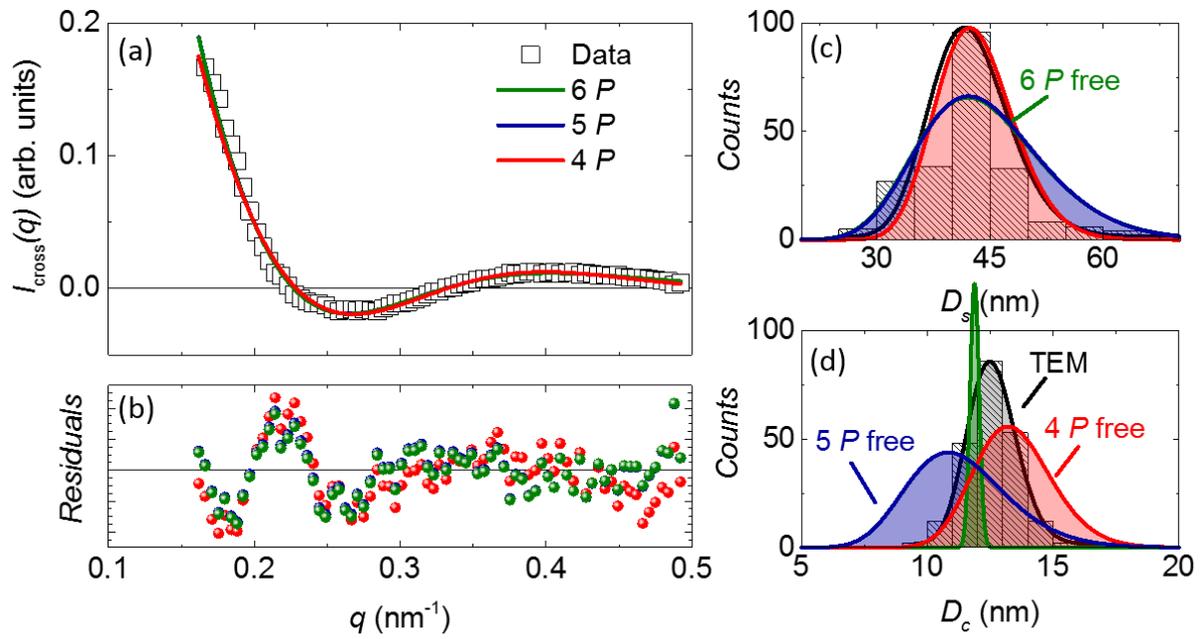

**Figure 3.** (a) Nuclear-magnetic cross term as a function of the scattering vector $q$ and for an applied magnetic field of $\mu_0 H_0 = 5$T (open squares). Solid lines: best fit performed with a varying number of free fit parameters $P$. (b) Corresponding residuals. (c) and (d) Log-normal distribution of the outer shell diameter (total particle size) and core size computed from the best-fitting parameters, respectively.

**Figure 4.** Corner plots of the fitting parameters obtained from the Bayesian refinements displayed in figure 3(a). The main diagonal displays the 1D probability distribution (histogram) for each fitting parameter. The off diagonal shows the cross correlations for each pair of parameters. In figure 4(a) all 6 parameters were allowed to vary; in (b) $\rho_s$ was fixed to 2.8 ×$10^{-6}$ Å$^{-2}$, whereas in (d), $\rho_s$ and $\sigma_s$ were fixed to 2.8 ×$10^{-6}$ Å$^{-2}$ and 0.12 (result from TEM analysis), respectively. Parameter units: $D_s$ and $D_c$ in nm, $\rho_s$ in $10^{-6}$ Å$^{-2}$, and $\rho_m$ in arbitrary units.

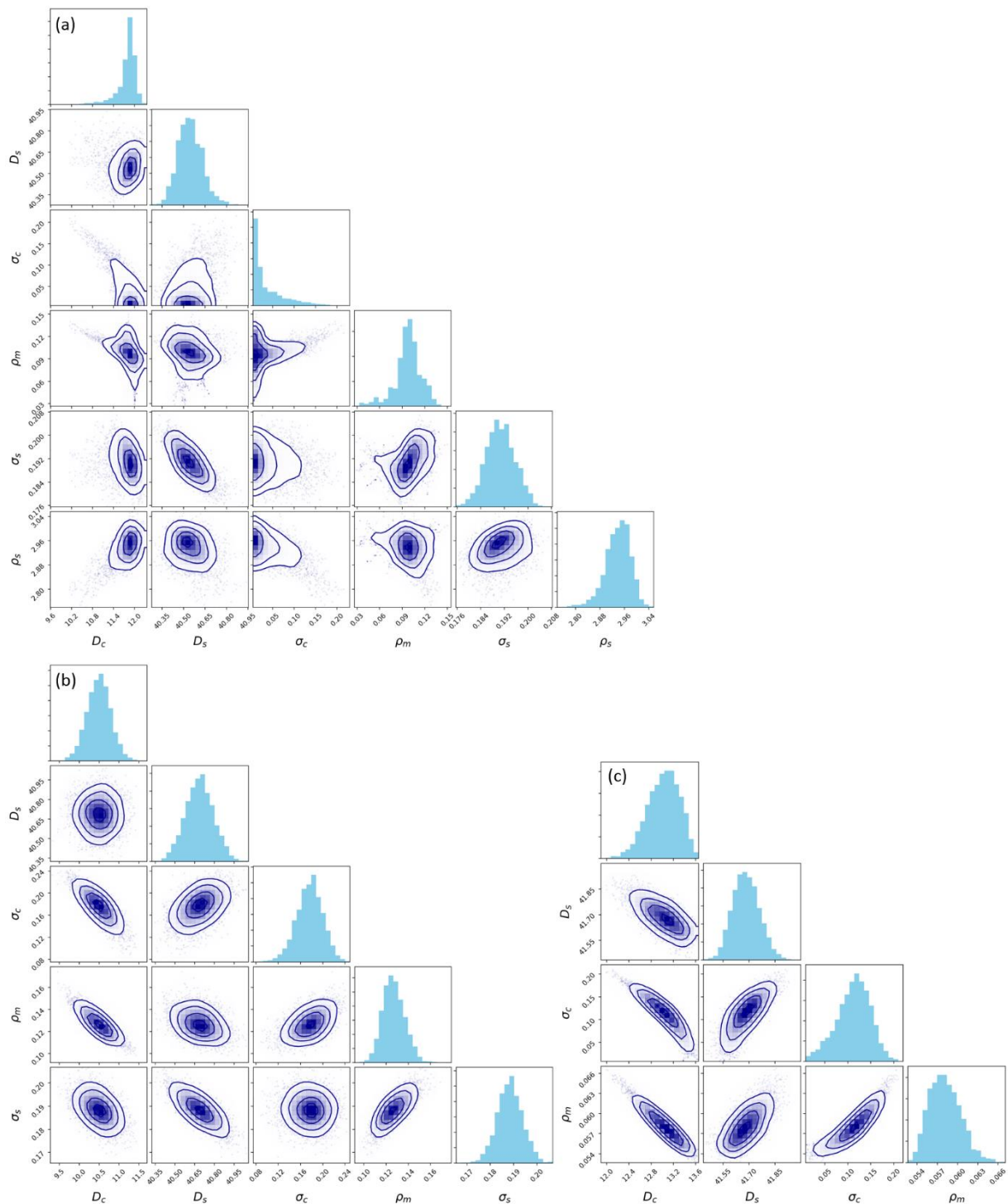

**Figure 5.** (a) Nuclear-magnetic cross term as a function of the scattering vector $q$ and for selected magnetic fields of $\mu_0 H_0$ = 0.02, 0.1, 1.0 and 5.0T. The best fits performed with 8 and 7 free parameters are sketched as dotted lines and solid lines, respectively. (b) Corresponding residuals. (c) and (d) Distributions of the outer shell diameter and core size, respectively. The log-normal distribution computed from the refinement performed with 8 and 7 free parameters $P$ are sketched by the green and red areas, respectively.

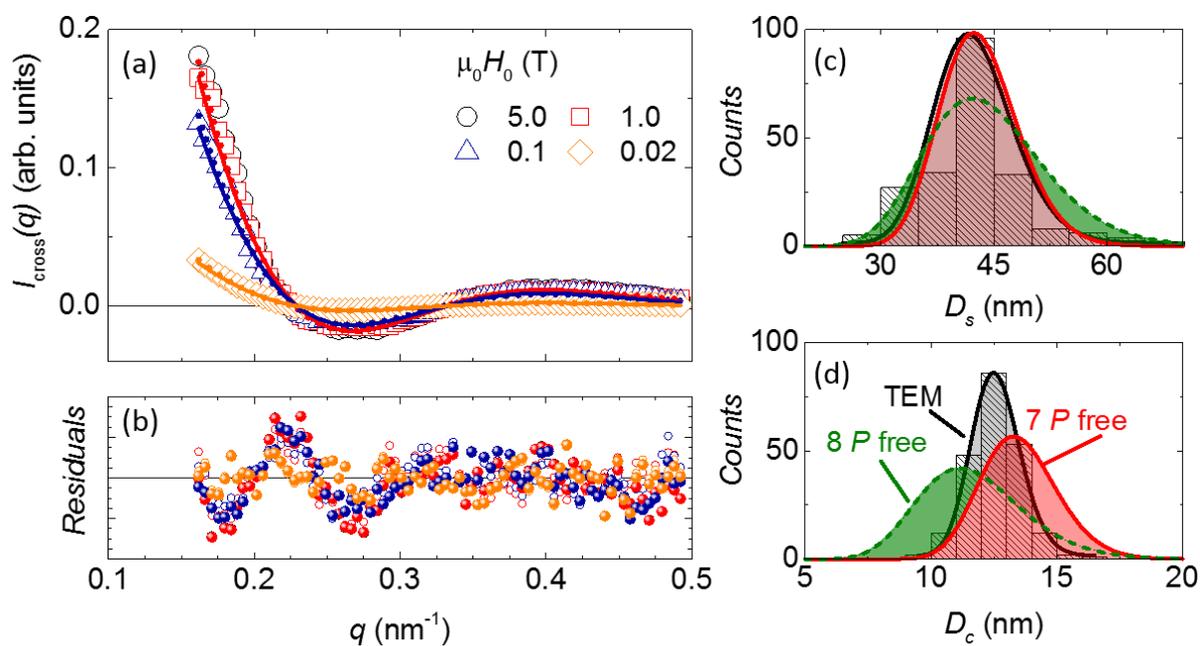

**Figure 6.** Corner plots of the fitted parameters obtained from the Bayesian refinement (global fit) displayed in figure 5(a). The main diagonal displays the 1D probability distribution (histogram) for each fitting parameter. The off diagonal two-dimensional plots show the cross correlations for each pair of parameters. In figure 6(a), $\rho_s$ was fixed to $2.8 \times 10^{-6}$ Å$^{-2}$ and in (b) $\rho_s$ and $\sigma_s$ were fixed to $2.8 \times 10^{-6}$ Å$^{-2}$ and 0.12 (result from TEM analysis), respectively. Parameter units: $D_s$ and $D_c$ in nm, $\rho_m$ in arbitrary units.

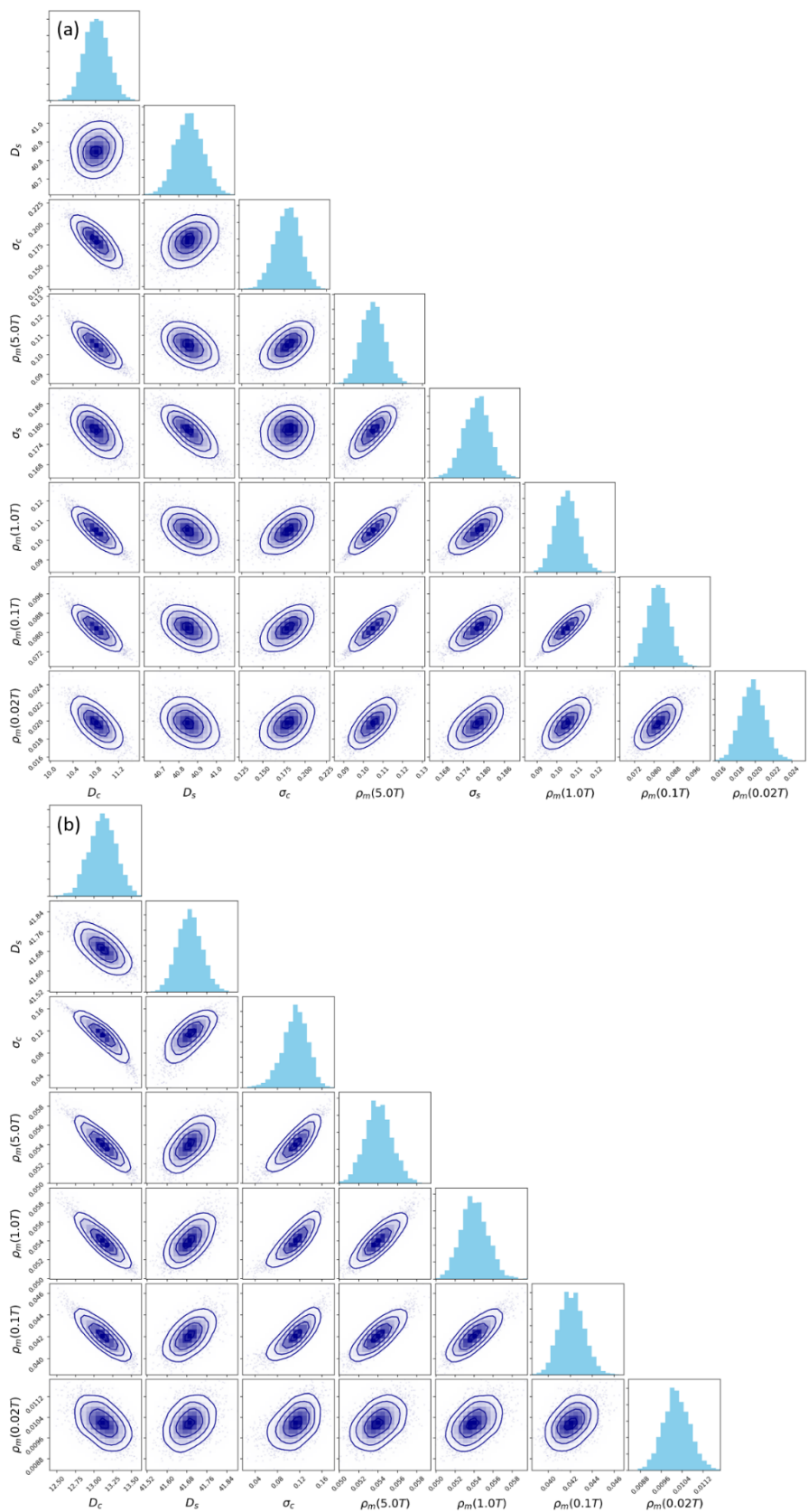

**Figure 7.** Log-linear plot of the positive branch of the magnetization curve (open squares) and best fit (solid red line) taken from figure 2(a). Yellow stars: $\rho_m$ values (rescaled using the values at 5 T for normalization) determined from the global fit of the SANS data set (figure 5(a)) and summarized in Table 3.

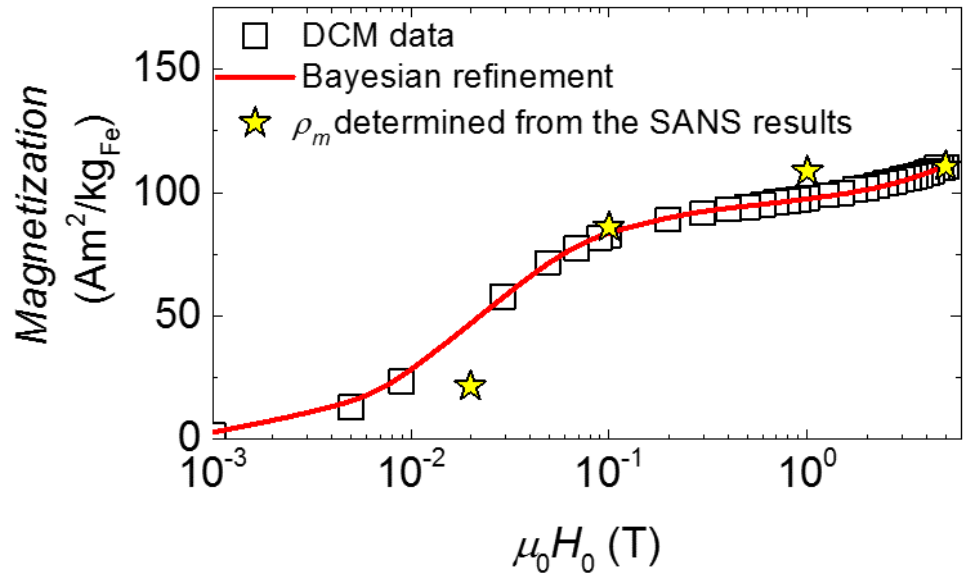

**Figure 8.** Global fit results of the isothermal magnetization curve plus the 4 field-dependent SANS curves. (a) Nuclear-magnetic cross terms. (b) Magnetization curve. The solid lines in (a) and (b) are the best-fit curves. For the global fit, the field values for the SANS measurements were fixed to 5, 1 and 0.1 T. The lowest field value had to be considered as a fit parameter to enable a good fit. $\rho_s$ was fixed to 2.8 ×10$^{-6}$ Å$^{-2}$. $\sigma_s$ and $\sigma_c$ were fixed to 0.12 and 0.08, respectively (result from TEM analysis). We also introduced a free parameter $c$ which corresponds to the normalization factor of the Langevin function between the SANS and DCM data. (c) and (d) are the corresponding residuals. (e) Corner plot of the fitted parameters obtained from the Bayesian refinement (global fit). The main diagonal displays the 1D probability distribution (histogram) for each fitting parameter. The off diagonal two-dimensional plots show the cross correlations for each pair of parameters. (f) and (g) show the distributions of the outer shell diameter and core size, respectively. The log-normal distributions computed from the refinement are indicated by the red area.

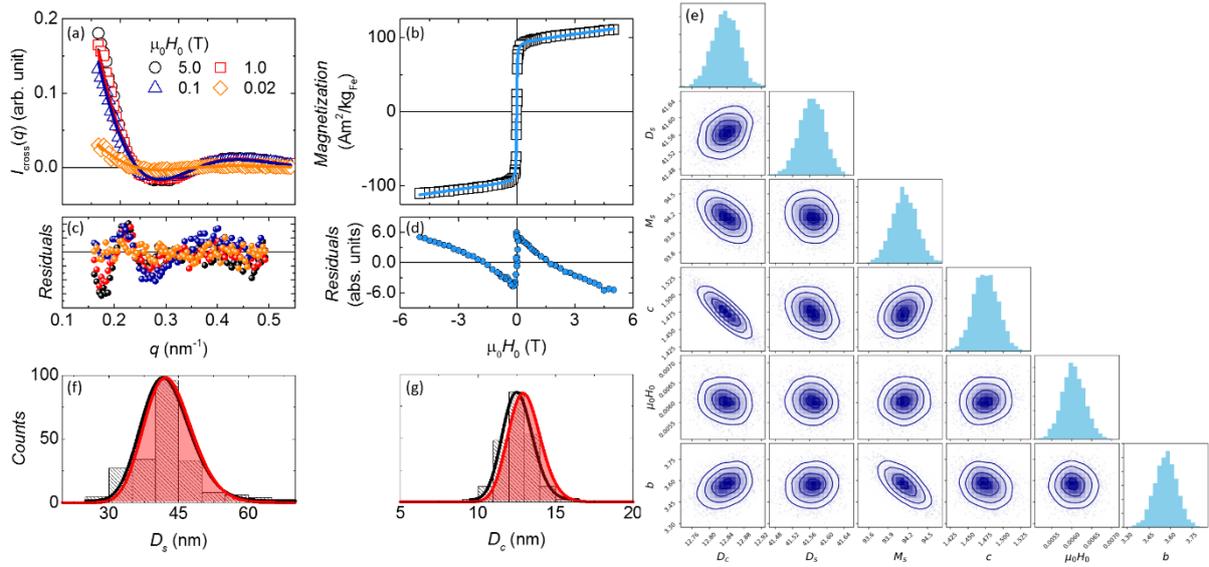

**Table 1.** Best-fitting parameters determined from the refinement of the TEM and DCM data.

| TABLE 1 STRUCTURAL AND MAGNETIC PARAMETERS | | | | | |
|---|---|---|---|---|---|
| | TEM data | Units | | DCM data | Units |
| $\langle D_c \rangle$ | $12.56 \pm 0.03$ | nm | $D_m$ | $12.1 \pm 0.1$ | nm |
| $\langle D_s \rangle$ | $42.5 \pm 0.2$ | nm | $M_s$ | $95.5 \pm 0.3$ | Am$^2$/kg$_{Fe}$ |
| $\sigma_c$ | $0.08 \pm 0.01$ | | $\sigma_m$ | $0.11 \pm 0.01$ | |
| $\sigma_s$ | $0.12 \pm 0.01$ | | $b$ | $3.2 \pm 0.1$ | J/kg$_{Fe}$ |

**Table 2.** Best-fitting parameters determined from the refinement of a single magnetic SANS data set.

| TABLE 2 STRUCTURAL AND MAGNETIC PARAMETERS | | | | |
|---|---|---|---|---|
| | 6 Parameters free | 5 Parameters free | 4 Parameters free | Units |
| $D_c$ | $11.8 \pm 0.2$ | $10.5 \pm 0.3$ | $13.0 \pm 0.3$ | nm |
| $D_s$ | $40.54 \pm 0.08$ | $40.7 \pm 0.1$ | $41.68 \pm 0.08$ | nm |
| $\sigma_c$ | $0.02 \pm 0.03$ | $0.18 \pm 0.02$ | $0.12 \pm 0.04$ | |
| $\rho_m$ | $0.10 \pm 0.01$ | $0.13 \pm 0.01$ | $0.058 \pm 0.002$ | arb. units |
| $\sigma_s$ | $0.191 \pm 0.005$ | $0.188 \pm 0.005$ | *fixed* (0.12) | |
| $\rho_s$ | $2.94 \pm 0.04$ | *fixed* (2.8) | *fixed* (2.8) | $10^{-6}$ Å$^{-2}$ |

**Table 3.** Best-fitting parameters determined from the Bayesian refinement of multiple, field-dependent magnetic SANS data sets with some shared (global) structural parameters.

| TABLE 3 STRUCTURAL AND MAGNETIC PARAMETERS | | | |
|---|---|---|---|
| | 8 Parameters free | 7 Parameters free | Units |
| $D_c$ | $10.8 \pm 0.2$ | $13.1 \pm 0.2$ | nm |
| $D_s$ | $40.86 \pm 0.07$ | $41.69 \pm 0.05$ | nm |
| $\sigma_c$ | $0.18 \pm 0.01$ | $0.11 \pm 0.02$ | |
| $\sigma_s$ | $0.178 \pm 0.004$ | *fixed* (0.12) | |
| $\rho_s$ | *fixed* (2.8) | *fixed* (2.8) | $10^{-6}$ Å$^{-2}$ |
| $\rho_m(5T)$ | $0.105 \pm 0.006$ | $0.054 \pm 0.001$ | arb. units |
| $\rho_m(1T)$ | $0.105 \pm 0.006$ | $0.053 \pm 0.001$ | ,, |
| $\rho_m(0.1T)$ | $0.082 \pm 0.005$ | $0.042 \pm 0.001$ | ,, |
| $\rho_m(0.02T)$ | $0.020 \pm 0.001$ | $0.0102 \pm 0.0005$ | ,, |

**Table 4.** Best-fitting parameters determined from the Bayesian refinement of multiple, field-dependent magnetic SANS data sets and DCM data with some shared (global) structural and magnetic parameters.

| TABLE 4 STRUCTURAL AND MAGNETIC PARAMETERS | | |
|---|---|---|
| | 6 Parameters free | Units |
| $D_c$ | $12.84 \pm 0.03$ | nm |
| $D_s$ | $41.57 \pm 0.03$ | nm |
| $M_s$ | $94.1 \pm 0.2$ | Am$^2$/kg$_{Fe}$ |
| $b$ | $3.57 \pm 0.08$ | J/kg$_{Fe}$ |
| $c$ | $1.48 \pm 0.02$ | |
| $\mu_0 H_0$ | $0.0060 \pm 0.0003$ | T |
| $\sigma_c$ | *fixed* (0.08) | |
| $\sigma_s$ | *fixed* (0.12) | |
| $\rho_s$ | *fixed* (2.8) | $10^{-6}$ Å$^{-2}$ |